\documentclass[reprint,amsmath,amssymb,aps,pra,showkeys,superscriptaddress,floatfix]{revtex4-2}

\usepackage[utf8]{inputenc}
\usepackage[english]{babel}
\addto\captionsenglish{}
\usepackage{graphicx}
\usepackage{color,epsfig}
\usepackage{array}
\usepackage{amsmath,amssymb}
\usepackage{natbib}
\usepackage{bbold}
\usepackage{dsfont}
\usepackage[normalem]{ulem}
\usepackage[numbered]{matlab-prettifier}
\usepackage[autostyle]{csquotes}
\usepackage{comment}
\usepackage{bm}
\usepackage{textcomp}
\usepackage{siunitx}
\begin{document}

\title{Microscopic magnetic field imaging with hot atoms via single-pixel imaging}

\author{Cyril Torre}
\email{cyril.torre@bristol.ac.uk}
\affiliation{Quantum Engineering Technology Labs, H. H. Wills Physics Laboratory and School of Electrical, Electronic, and Mechanical Engineering, University of Bristol, BS8 1FD, United Kingdom.}%
\author{Jordan Brass}
\affiliation{School of Electrical, Electronic, \& Mechanical Engineering, University of Bristol, Tyndall Avenue, BS8 1FD, United Kingdom.}%
\author{Sebastien Bisdee}
\affiliation{Quantum Engineering Technology Labs, H. H. Wills Physics Laboratory and School of Electrical, Electronic, and Mechanical Engineering, University of Bristol, BS8 1FD, United Kingdom.}%
\affiliation{Quantum Information Science and Technologies Centre for Doctoral Training, H. H. Wills Physics Laboratory and School of Electrical, Electronic, \& Mechanical Engineering, University of Bristol, BS8 1FD, United Kingdom}%
\author{Mohammed T. Rasheed}
\affiliation{H.H. Wills Physics Laboratory, University of Bristol, Tyndall Avenue, BS8 1FD, United Kingdom.}%
\author{Giacomo Ferranti}
\affiliation{Quantum Engineering Technology Labs, H. H. Wills Physics Laboratory and School of Electrical, Electronic, and Mechanical Engineering, University of Bristol, BS8 1FD, United Kingdom.}%
\author{Carrie A. Weidner}
\affiliation{Quantum Engineering Technology Labs, H. H. Wills Physics Laboratory and School of Electrical, Electronic, and Mechanical Engineering, University of Bristol, BS8 1FD, United Kingdom.}%

\date{\today}

\begin{abstract}
    In recent years, sensors based on hot atomic vapor cells have emerged as a compact and highly sensitive means of measuring magnetic fields. Such sensors have been deployed in the field for the measurement of, e.g. biological systems, representing a promising practical application of quantum technologies. However, it remains challenging to obtain high-resolution magnetic field images from these sensors, and in most cases the spatial resolution of the system is limited by the sensor size. Here, we demonstrate the combination of single-pixel imaging (SPI) techniques with an atom vapor Faraday magnetometer to achieve microscopic magnetic field imaging. We demonstrate magnetic field imaging with a spatial resolution of $\approx~\SI{62.5}{\micro\meter}$, limited only by the resolution of our DMD projection system and the absence of magnetic shielding in our experimental setup.
\end{abstract}


\maketitle

\section{\label{sec:intro}Introduction}
Atomic sensors have proven to be among the most sensitive for measuring magnetic fields, achieving subfemtotesla sensitivity~\cite{Kominis2003,Dang2010,Ledbetter2008}. High-resolution, high-sensitivity imaging of magnetic fields is thus a logical next step in the development of this technology. 
A demonstration of magnetic field imaging using ultracold atoms achieved a spatial resolution of $\SI{3}{\micro\meter}$ and a sensitivity of $\SI{300}{\pico\tesla}$~\cite{BEC_MfieldImaging}. However, cold atom systems require relatively complex, bulky experimental systems, and compactification of such systems into something suitable for practical applications remains challenging. 
In addition, one must trap the cold atoms close enough to the sample in order to sense the magnetic field, rendering biological applications particularly challenging. 
With the recent emergence of hot atomic vapor cells, the experimental apparatus can be dramatically simplified while maintaining high sensitivity up to a few $\SI{}{\femto\tesla\per\sqrt{\hertz}}$ with a miniature millimetre-scale Rubidium (Rb) vapor cell~\cite{Kitching2007,Compact_mag, Griffith2010}. Vector magnetometry with $\SI{}{\pico\tesla\per\sqrt{\hertz}}$ sensitivities have been demonstrated in hot atom vapor~\cite{Fernholz2019}. Such systems have also been made compact and portable, demonstrating magnetic field sensing from biological sources, such as the brain~\cite{Brookes2022}, the heart~\cite{Dawson2023} or the eyes~\cite{Westner2021}.

However, demonstrations of magnetic field imaging using hot atoms have been, to-date, limited. A CCD camera has been used to image atoms and detect magnetic field variations per $10\times\SI{10}{\micro\meter}$ pixel~\cite{Narducci2009}. Arrays of microfabricated optically-pumped magnetometers (OPMs) have been used for highly sensitive (tens of \SI{}{\femto\tesla\per\sqrt{\hertz}}) magnetic field imaging~\cite{Knappe2017}, but their resolution is fundamentally limited by the magnetometer system size. Other methods produced images by scanning a $\SI{100}{\micro\meter}$-scale beam across a vapor cell~\cite{Riis2023}. Microwave field imaging with micrometer spatial resolution has been achieved in alkali vapor~\cite{Treutlein2012,Treutlein2015,Prajapati2024}; this method was also extended to DC field imaging~\cite{Mileti2015}.

In this work, we propose combining vapor cell magnetometry with single-pixel imaging (SPI) techniques~\cite{Duarte2008}. In particular, as our measurement is the polarization rotation of light passing through the cell via the Faraday effect~\cite{Schatz1969}, we implement SPI polarimetry~\cite{Lancis2012,Padgett2015,Zhao2020,Foreman2020,Wang2022} as a means of generating the magnetic field images. A detailed review of SPI can be found in Ref.~\cite{Gibson2020}. The SPI technique involves measuring the transmitted light intensity through the sample for a series of known patterns projected by a spatial light modulator, usually a digital micromirror device (DMD). The resulting transmitted intensities, recorded on a photodetector, are then weighted according to the corresponding DMD patterns to reconstruct the image~\cite{Gibson2020}. A key advantage of this method is that DMDs are relatively compact and can be easily integrated to most hot atom vapour magnetometer technologies. Furthermore, photodetector technology is well-established, offering a high quantum efficiency and low noise floor, making this system amenable to eventual use with light sources operating below the shot noise limit~\cite{Wolfgramm2010,Horrom2012,Bai2021}.

This paper is organized as follows: We present relevant theory in Sec.~\ref{sec:theory} before describing the experiment in Sec.~\ref{sec:exp}. Results follow in Sec.~\ref{sec:results}, and Sec.~\ref{sec:conc} concludes.

\section{\label{sec:theory}Theory}

Our chosen basis for the implementation of SPI is the set of differential Hadamard patterns, as they offer a higher signal-to-noise ratio performance compared to other patterns~\cite{SPIFvsH}. These are defined by a series of Hadamard matrices, where the Hadamard matrix of dimension two can be defined by
\begin{equation}
     H_2 = \left(\begin{matrix}
        1 & 1 \\
        1 & -1
    \end{matrix}\right).
\end{equation}
To reconstruct an image with a spatial resolution of $n \times n$ pixels, all patterns can be generated using a single Hadamard matrix of dimension $n^2$, where:
\begin{equation}
     H_{n^2} = \left(\begin{matrix}
         H_{n^2/2} & H_{n^2/2} \\
         H_{n^2/2} & -H_{n^2/2}
    \end{matrix} \right) =  H_2\otimes H_{n^2/2},
\end{equation}
and $n$ is a power of $2$. Each pattern corresponds to $H_{n^2}$, a single row of dimension $1 \times n^2$, to be afterward reshaped into a $n \times n$ binary matrix and to be loaded into the DMD. Because the DMD cannot project a negative intensity, each pattern needs to be projected twice, first by projecting the positive value as $1$ and negative as $0$, and vice versa. Hadamard matrices are unitary and invertible, forming an orthogonal basis. Therefore, an image $s_{n \times n}$ can be easily mathematically reconstructed from the measured intensities of each pattern, $I_\pm$, which are vectors of dimension $n^2 \times 1$ corresponding to the positive and negative patterns, respectively. Hence, we have:
\begin{equation}
    s_{n\times n} = (H_{n^2}^+ - H_{n^2}^-)^{-1} .(I_+ - I_-),
\end{equation}
where $s_{n \times n}$ is a vector of dimension $n^2 \times 1$ that needs to be reshaped into an $n \times n$ matrix, representing the reconstructed image of the sample.

The magnetic field is sensed via the Faraday effect, also known as Faraday rotation (FR)~\cite{Schatz1969}. This phenomenon is described by a rotation of the incident linear polarization transmitted through a medium in the presence of a magnetic field. Mathematically, we have
\begin{equation}
    \Phi = B L v,
    \label{eq:FR}
\end{equation} 
where $\Phi$ is the rotation of the polarization of the input light, $B$ the magnetic field, $L$ the length of the medium (here the vapor cell), and $v$ the Verdet constant, which is a wavelength-dependent optical property of a given medium. The Verdet constant needs to be calibrated prior to any magnetic field sensing. It sets the sensitivity of the magnetometer, but is also sensitive to temperature, frequency of the probe, and the external magnetic field~\cite{Bai2021}. Indeed, if the magnetic field is too strong, the energy levels of the rubidium atoms in the vapor cell will shift, leading to a change in the Verdet constant. Therefore in this work, Helmholtz coils are used to add a bias on the magnetic field surrounding the cell in order to zero the mean field produced by the sample and ensure the same sensitivity within the range of the magnetic field produced by the sample.

\section{\label{sec:exp}Experimental Setup}
\subsection{Experiment description}
\begin{figure}[!ht]
    \centering
    \includegraphics[width=\columnwidth]{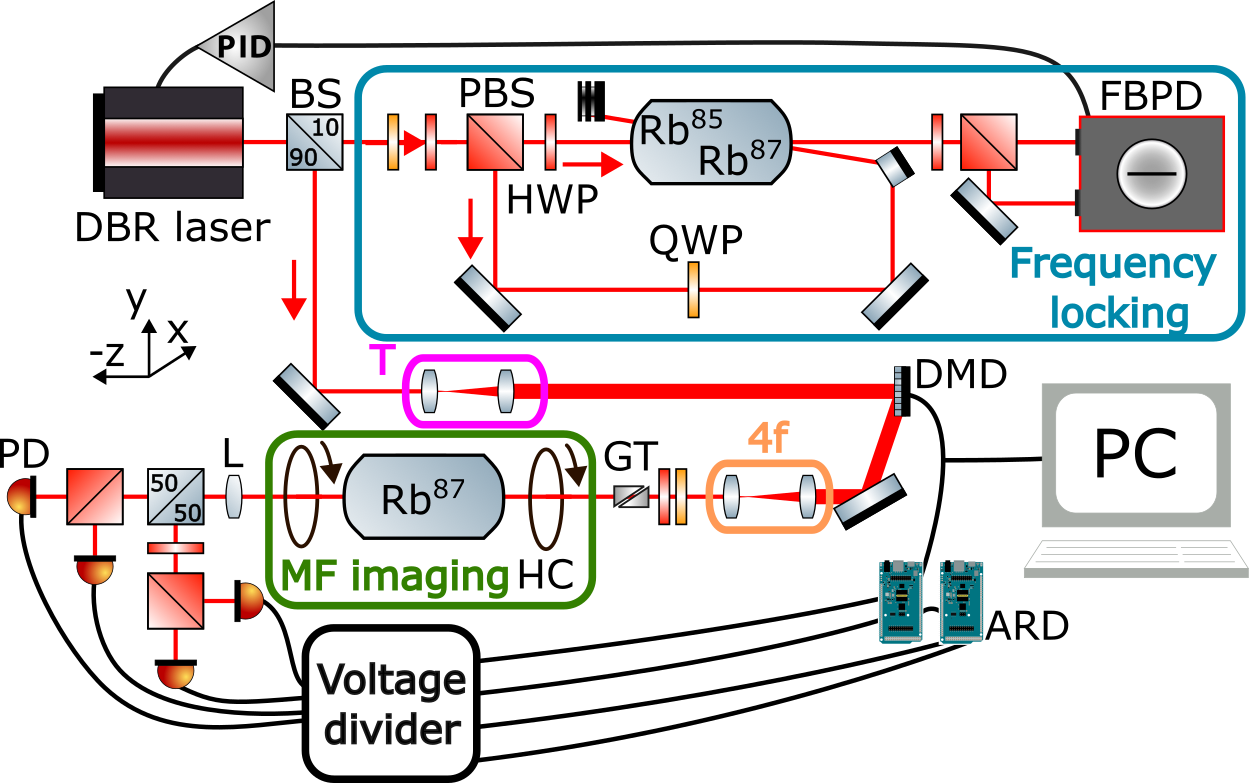}
    \caption{\textbf{Microscopic magnetic field imaging experimental setup.} A DBR laser (\textit{DBR795PN}, \textit{Thorlabs}) is split by a $90:10$ beamsplitter (BS), where $10~\%$ of the light is sent to a frequency locking setup (blue rectangle) with a PID (\textit{Moku:GO}, \textit{Liquid Instrument}) used to lock the laser on an atomic transition $^{87}$Rb, $F=2$ to $F=2'$ on the $D1$ line. Then, $90~\%$ of the light is sent toward the DMD to perform single-pixel imaging. The two lenses in the optical telescope $T$ increase the beam's size by $20$ to improve the uniformity of the beam striking the DMD surface. After the DMD, the light is transmitted toward a second cell filled with $^{87}$Rb, where a $4f$ system is used to image the centre of the cell and the magnetic field is measured via the Faraday effect. Helmholtz coils (HC) apply a uniform and constant magnetic field to the cell. Finally, a non-polarizing beamsplitter (BS) splits the beam between two polarizing beamsplitters (PBS) with one preceded by a half-wave plate (HWP) at $\SI{22.5}{\degree}$, enabling intensity measurements on the horizontal/vertical and diagonal/anti-diagonal bases with photodetectors PD (\textit{PD100A2}, \textit{Thorlabs}). The DMD is triggered by one of two Arduino Giga R1 (ARD) microcontrollers, while the photocurrents are acquired by both Arduino microcontrollers.}
    \label{fig:experiment}
\end{figure}

The experimental setup is illustrated in Fig.~\ref{fig:experiment}. A $\SI{795}{\nano\meter}$ pigtailed DBR laser (\textit{DBR795PN}, \textit{Thorlabs}) is first split by a $90:10$ beamsplitter (BS). The $10\%$ of light that is picked off is used to lock the laser onto the $F=2$ to $F'=2$ transition of the $D1$ line of $^{87}$Rb via polarization spectroscopy locking~\cite{freq_locking} (see Appendix~\ref{app:spec} for further details). 
The remaining $90~\%$ of the laser power is directed toward the magnetic field imaging system. An optical telescope (with lenses of focal length $\SI{12.5}{\mm}$ and $\SI{200}{\mm}$) expands the beam size to ensure a nearly uniform intensity across the region illuminated by the DMD (\textit{DLP650LNIR}, \textit{Vialux}) pattern. The magnified beam is then reflected by the DMD toward a vapor cell containing $^{87}$Rb atoms. The temperature of the cell is stabilized at $\SI{70}{\celsius}$ using a glass cell heater (\textit{GCH25-75}, \textit{Thorlabs}) and a temperature controller (\textit{TC300B}, \textit{Thorlabs}). It is worth noting that the cell is not magnetically shielded from its environment.

Immediately before the vapor cell, a $4f$ imaging system (with two lenses of focal length of $\SI{250}{\mm}$ and $\SI{200}{\mm}$) is used to collimate the beam along the cell axis. To maximize the power incident on the cell and ensure linear polarization, a combination of a quarter-waveplate (QWP), half waveplate (HWP), and Glan-Taylor polarizer (GT) are employed, reaching an input light power of $\approx \SI{800}{\micro\watt}$. A pair of coils in a Helmholtz configuration is used to apply a uniform DC magnetic field along the direction of light propagation through the cell; this magnetic field acts as a bias. Finally, the output light ($\approx \SI{400}{\micro\watt}$ after locking the laser) is first focused by a lens (L) with a focal length of $\SI{125}{\mm}$ and then split by a $50:50$ BS to direct the light toward the sets of photodiodes used to make the polarimeter. The transmitted beam is first directed to a polarizing BS (PBS) to measure the horizontal (H) and vertical (V) polarization components, while the reflected light is sent to a second PBS, preceded by a HWP with its fast axis oriented $\SI{22.5}{\degree}$, enabling the measurements of the diagonal (D) and anti-diagonal (A) polarization components. Each photodetector (\textit{PD100A2}, \textit{Thorlabs}) has its gain set at $10^3$ to maximise the signal to noise ratio while also avoiding saturation. The photocurrent passes through a voltage divider that allows us to make use of the full dynamic range of the microcontrollers and is monitored by two Arduino giga R1 boards. Each Arduino is a $16$-bit ADC, with a sampling rate approaching $\SI{}{\mega\hertz}$ through the use of the STMSpeeduino library ~\cite{Arduino_lib}.

\subsection{Data acquisition}

The data acquisition is done by starting the pre-loaded sequence on the DMD. For each pattern, the DMD will send a trigger to the Arduino boards, after which the Arduino boards start to collect data from the photodiodes after a delay of $\SI{70}{\micro\second}$ to avoid photocurrent fluctuation due to the initialization of the DMD's mirrors. Then, $\SI{100}{\micro\second}$ before the end of the illumination pattern, the trigger falls, stopping the data acquisition. The illumination time (time for which a pattern is projected by the DMD) is $\SI{600}{\micro\second}$ long, while the picture time (time between two successive patterns) is set to $\SI{6}{\milli\second}$. This delay is set to allow enough time for the Arduino boards to send the data through the serial line after each pattern before the start of the next one. While the DMD can be faster (up to $\approx \SI{20}{\kilo\hertz}$), the imaging speed in this work is limited by the data transfer. 
Finally, in this work, we report an average of $\approx 220$ detections (i.e., one measurement in the H/V and D/A basis) for each pattern.

\section{\label{sec:results}Experimental results}
\subsection{Spatial resolution}
\begin{figure}[!ht]
    \centering
    \includegraphics[width=\columnwidth]{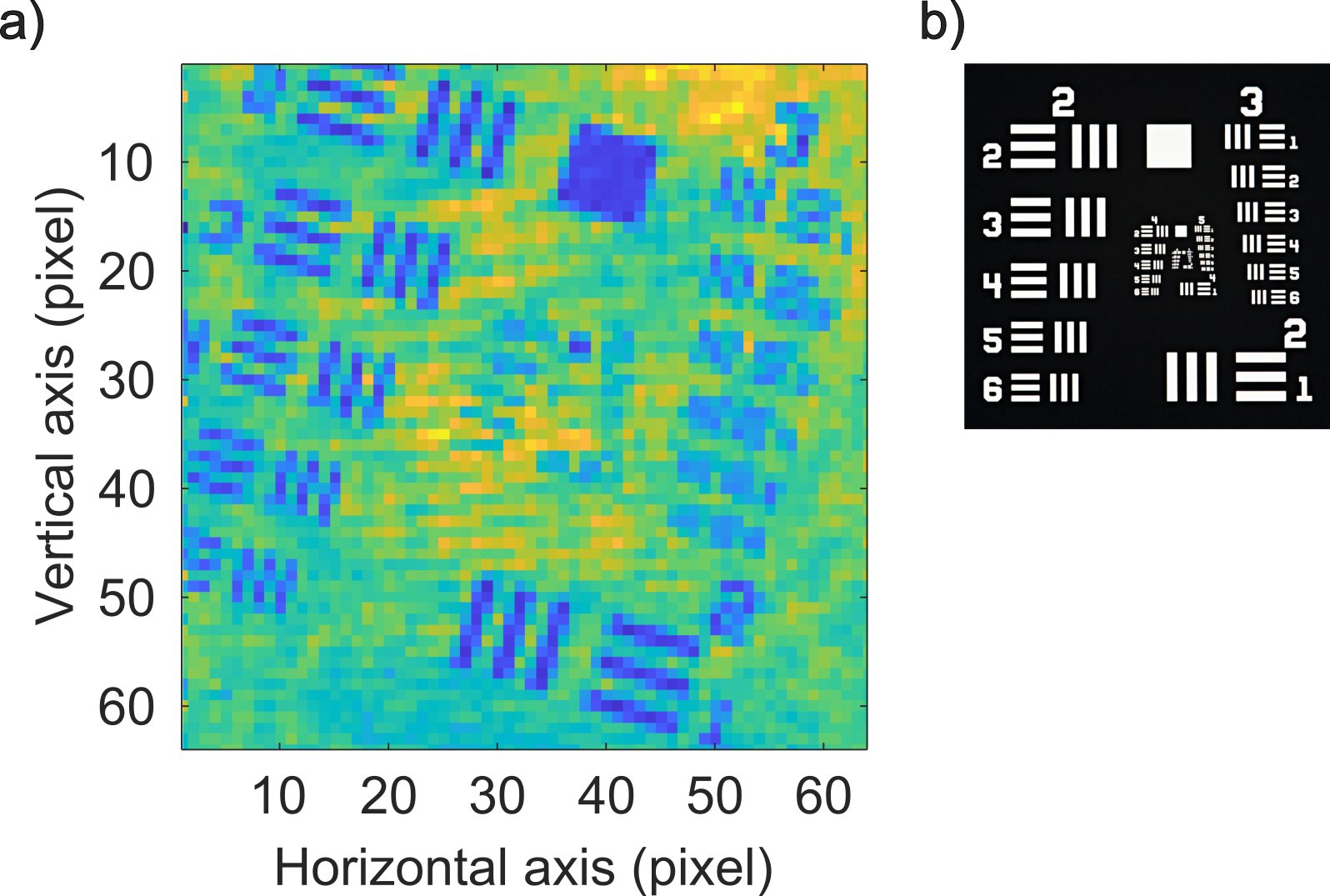}
    \caption{\textbf{Single pixel imaging of a resolution sample target  (\textit{R1DS1P}, \textit{Thorlabs}).} (a) Single-pixel imaging using differential Hadamard patterns for $n=64$. (b) Image of the 1951 USAF test target. On (a), we can distinguish the line pair from group 2, element 6 (bottom left on the image) while the element 1 from group 3 is not perfectly distinguishable. This gives a pixel size between $\SI{62.5}{\micro\meter}$ and $\SI{70}{\micro\meter}$.}
    \label{fig:Res64}
\end{figure}

Each photodetector measures the intensity of each linear polarization component (horizontal, vertical, diagonal and anti-diagonal) with a spatial resolution determined by the size of the Hadamard patterns for a given beam diameter. For $n= 64$, the spatial resolution (here the actual size of a pixel in the reconstructed image) from our SPI setup has been experimentally characterised by imaging a 1951 USAF resolution test target, as shown in Fig.~\ref{fig:Res64}. From the figure, we can distinguish the line pair from group 2, element 6 (bottom left on the image) while the element 1 from group 3 is not perfectly distinguishable. Based on the line pair dimensions provided by the manufacturer, we can estimate the pixel size in our image. This provides a spatial resolution between $\SI{62.5}{\micro\meter}$ and $\SI{70}{\micro\meter}$, corresponding to an illumination area of approximately $(\SI{4}{}\times\SI{4
}{})~\SI{}{\mm^2}$ from the probe beam.

\subsection{Calibration of the Verdet constant}
\begin{figure}[!ht]
    \centering
    \includegraphics[width=\columnwidth]{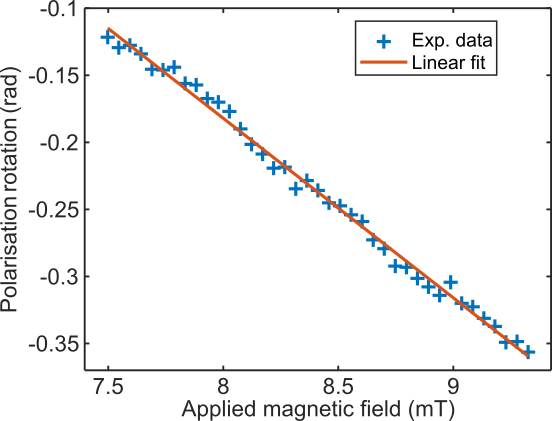}
    \caption{\textbf{Calibration of the Verdet constant.} Polarization rotation of a single DMD pixel as a function of the applied magnetic field. The blue crosses are experimental datapoints, while the orange trace is a fit using Eq.~\eqref{eq:FR} with free parameter \textit{v}, and an adjusted R-square $0.994$. The Verdet constant is estimated at $(1.91 \pm 0.05)\times 10^3~\SI{}{\radian\per\tesla\per\meter}$.}
    \label{fig:Vfit}
\end{figure}

Prior to performing magnetic field imaging with the hot vapor cell via the Faraday effect described in Eq.~\eqref{eq:FR}, the Verdet constant $v$ needs to be estimated. 
In this work the Verdet constant is estimated by measuring the polarization rotation for a single pixel (ideally at the centre of the image) while the axial magnetic field is swept with a pair of coils in a Helmholtz configuration in the presence of the sample. The magnetic field generated by the coils is measured with a gaussmeter (\textit{MF100}, \textit{FLIR Extech}), at a single point between the two coils, equally distant from both. Because the diameter of the coils ($\SI{19.5}{\centi\metre}$) is large compared to the dimensions of the vapor cell (a cylinder $\SI{70}{\mm}$ long with a diameter of $\SI{25}{\mm}$), we can assume that a single point measurement is sufficient as the magnetic field should be fairly uniform in the area covered by the cell. The results are shown in Fig.~\ref{fig:Vfit}, and the Verdet constant is estimated at $(1.91 \pm 0.05)\times 10^3~\SI{}{\radian\per\tesla\per\meter}$ for our cell, 
which is of the same order of magnitude reported in previous work~\cite{Bai2021}. 

\subsection{Magnetic field imaging}
The sample consists of $10$ rectangular permanent magnets with individual size of $(38.1\times6.4\times 6.4)~\SI{}{\milli\meter\cubed}$ (axially magnetised, \textit{Part:9168},\textit{Radial Magnets}). They are parallel to the vapor cell, with their north pole pointing toward the positive $z$ axis. Hence, a magnetic field is formed, decreasing along the horizontal $x$ axis, where we can assume a uniform magnetic field alongside the $y$ axis over the $\SI{4}{\milli\meter}$ of the magnetic field image.

The polarization rotation is estimated by monitoring the intensities of each linear polarization component and is given by:
\begin{equation}
    \Phi = \frac{1}{2} \arctan\left(\frac{\overline{D-A}}{\overline{H-V}}\right),
    \label{eq:Phi}
\end{equation}
where $H, V, D, A$ are the intensities of the four linear polarization components (horizontal, vertical, diagonal and anti-diagonal) and with $\overline{\alpha-\beta}$ (for $\alpha = H,D$ and $\beta = V, A$) the normalized polarization within their corresponding basis given by
\begin{equation}
    \overline{\alpha-\beta} = \frac{\alpha-\beta}{\alpha+\beta}.
\end{equation}

\begin{figure}[!ht]
    \centering
    \includegraphics[width=\columnwidth]{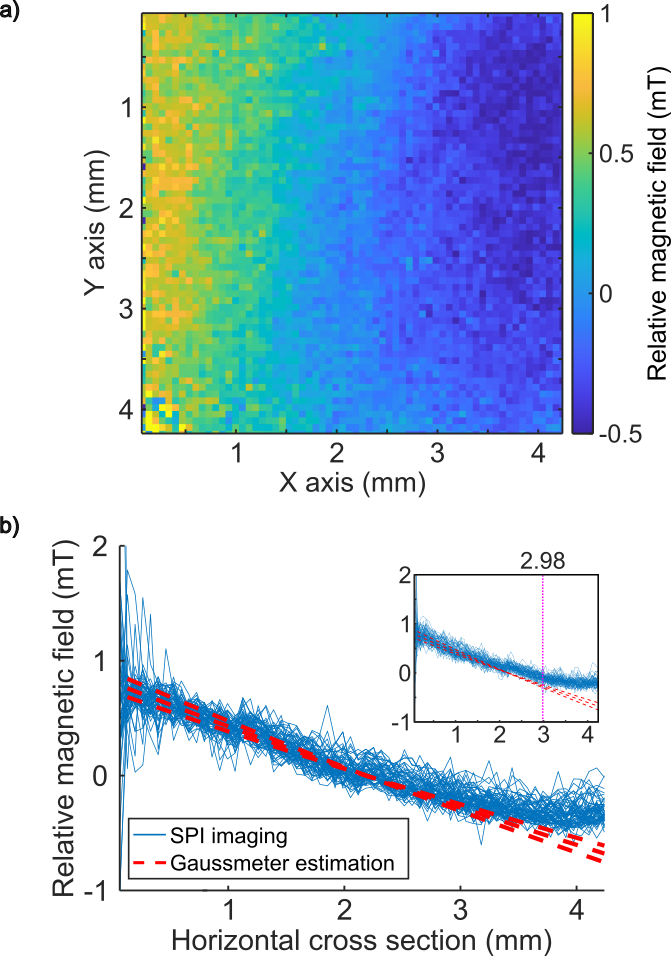}
    \caption{\textbf{Microscopic magnetic field imaging.} (a) Magnetic field imaging with a spatial resolution of $\approx\SI{62.5}{\micro\meter}$ integrated over the vapor cell from a magnetic sample made of permanent magnets on the side. The image is done by performing differential Hadamard patterns with $n = 64$. (b) Horizontal cross sections of the image in (a). The blue traces are the individual cross sections (taken per pixel along the $y$-axis) while the red dashed lines correspond to the magnetic field gradient estimated by the sample calibration (see Appendix~\ref{Appendix:CalibSample} for further details). The inset corresponds to a cross section of magnetic field SPI with a higher bias magnetic field given by the Helmholtz coils and showing a tail-off occurring sooner on the horizontal axis.}
    \label{fig:SPI_MF}
\end{figure}

Fig.~\ref{fig:SPI_MF}(a) demonstrates microscopic magnetic field imaging after calibration of the Verdet constant, showing the expected linear gradient across the $x$ axis. Fig.~\ref{fig:SPI_MF}(b) illustrates the horizontal cross-sections of Fig.~\ref{fig:SPI_MF}(a). To compare our measurements with a reference, the magnetic field of the $z$ component from the sample has been estimated by a gaussmeter (see Appendix~\ref{Appendix:CalibSample} for more details). While the gradient measured by the reference alongside the horizontal axis is precise, as the displacement between the gaussmeter and the sample is done with a motorized translation stage during the calibration process, the overall distance between the sample and the probe beam has been estimated with a ruler. Hence, we assume a lack of accuracy of $\pm \SI{2}{\milli\meter}$ over the distance between the sample and the probe beam along the horizontal axis, which is represented by plotting three gradients from the gaussmeter estimation on Fig.~\ref{fig:SPI_MF}(b). From this figure, we can see that the measured magnetic field from the SPI setup is in agreement with the gaussmeter estimation.

However, we can note that measured field gradient is no longer constant after $x=\SI{3.5}{\milli\meter}$. This arises due to the limited dynamic range of our system. We carefully choose the external bias field (equivalently, the laser frequency) to operate in a regime where the polarization rotation changes linearly with applied field due to the sample; outside of this regime, the Verdet constant changes and nonlinearities arise. Indeed, we observe that increasing the bias can shift this linear regime such that the nonlinear effect occurs closer to the origin of the horizontal axis, around $x=\SI{2.98}{\milli\meter}$ (see the inset of Fig.~\ref{fig:SPI_MF}(b)); decreasing the shift will cause nonlinearities to arise around $x = 0$. This could be avoided by working with a sample with a weaker magnetic field; here we use a relatively strong field to demonstrate the proof-of-principle of our method.

\section{\label{sec:conc}Discussion and conclusion}

We have successfully demonstrated magnetic field imaging with a microscopic spatial resolution of $\approx \SI{62.5}{\micro\meter}$, where the magnetic field is sensed via the Faraday effect of a hot vapor cell and the imaging uses single-pixel polarimetric methods. This represents the first demonstration of magnetic field SPI using a compact warm atomic vapor setup, and in what follows, we will outline a set of improvements that can be straightforwardly implemented in future work, with an eye towards higher sensitivities, compactification, and improved resolution.

First, while our magnetometer currently measures fields in the $\SI{}{\milli\tesla}$ range, this can be dramatically improved through the use of a magnetic shield that eliminates the effects of background fields, as demonstrated in Refs.~\cite{Wolfgramm2010,Horrom2012,Bai2021}. To avoid the nonlinear effects due to variations in the Verdet constant, one can also decrease the imaging area to a region where the Verdet constant does not vary appreciably.

We further note that the resolution of our system is limited only by the DMD projection system optics and the available laser power (which limits us to images with a $64\times 64$ pixel size); improvements in these would improve the resolution of our magnetometer, and we are far from any fundamental limits on resolution. While we do not address SPI for different magnetometry techniques involving hot atomic vapor cells (such as optically-pumped magnetometry~\cite{Alexandrov2003-th,Tierney2019}), we can see direct application of SPI techniques onto these different magnetometers, allowing for magnetic field imaging up to and beyond the resolution limit set by the DMD itself~\cite{Dai2018}.

The experimental setup could be easily integrated into a compact system as demonstrated in Ref.~\cite{Dawson2023}. In this work, as in ours, differential Hadamard patterns have been chosen due to the higher signal-to-noise ratio that this method offer compare to other~\cite{SPIFvsH}. However, to take an image of $n \times n$ pixels, in this work, $2 \times n^2$ patterns are required. Therefore, the imaging time does not scale very well when a higher resolution is required. However, compressive methods have demonstrated high quality imaging using a smaller the number of patterns~\cite{Radwell2017}. We do not explore this in this work, but similar methods can be employed in our scheme.

The main limitation in the data acquisition speed was due to the serial line between the Arduino boards and the computer. Indeed, while the acquisition time was set to $\approx \SI{400}{\micro\second}$ for each pattern, the spacing between each pattern was set to $\SI{6}{\milli\second}$, to allow enough time to transfer the data from the Arduino boards to the computer between each pattern. Writing the data onto the SDRAM on a memory card straight away on the Arduino boards can speed up the imaging process, as would the use of improved hardware. The fundamental speed limit of SPI is limited by the current speed of the DMD, which is typically in the $\approx 10-\SI{20}{\kilo\hertz}$ range for most commercially-available devices, although methods for accelerated pattern projection rates in the tens of MHz have been demonstrated~\cite{Liang2022}.

Finally, this method of imaging also facilitates the measurement of quantum probes via homodyne detection, where squeezed vacuum states have previously been used to enhance the sensitivity of hot atom vapor magnetometers~\cite{Wolfgramm2010,Horrom2012,Bai2021}. In future work, we will apply this scheme with a squeezed-vacuum source derived from the light-atom interaction, such as the polarisation-self-rotation (PSR) effect~\cite{Mikhailov2008}. While magnetometry has been demonstated with the more typical nonlinear-crystal-derived squeezing methods~\cite{Wolfgramm2010}, the advantages of PSR-based methods source compared to OPO/OPA cavities~\cite{Wen2017} include its compactness (a single beam transmitted through a vapor cell), energy efficiency (eliminating the need for SHG conversion), and the use of self-homodyne detection, which avoids the need for quantum-noise locking schemes~\cite{Zhang2017}. A recent demonstration of high stability squeezing enabled by AI control~\cite{Zhao2025} demonstrated robust PSR squeezing levels of $\SI{4.3}{\decibel}$. 
Similar methods would allow us to reach higher sensitivity and smaller variances which can lead to higher imaging contrast beyond the shot-noise limit~\cite{Taylor2013,SabinesChesterking2019,Atkinson2021}. The SPI method is well-suited for measurements based on squeezed light, as it only requires two balanced photodetectors instead of an array of matched and balanced camera pixels. Furthermore, work has shown progress towards real-time quantum-enhanced imaging using SPI~\cite{Pooser2013}. Thus, the combination of the two methods would lead to a compact, high sensitivity quantum-enhanced magnetometer suitable for day-to-day application.



\section*{Acknowledgment.}
The authors acknowledge funding via an Accelerated Development Fund from the EPSRC Quantic Hub for Quantum Imaging and Sensing, EP/T00097X/1. C.T. further acknowledges funding from a Quantic Doctoral. The authors would like to thank Dr Rowan Hoggarth, Dr. Shan Jiang, Harry Kendell, and Dr Vineet Bharti for their useful discussions.

\appendix
\section{\label{app:spec}Atomic spectroscopy frequency locking}
In this work, the frequency of the laser is locked using spectroscopy frequency locking~\cite{freq_locking}. Before locking the laser to an atomic transition, we first ensure that the hyperfine levels of the $D_1$ line can be detected by the balanced photodetectors (\textit{PDB210A/M}, \textit{Thorlabs}). To control the power distribution between the pump (reflected) and probe (transmitted) beams, we use a quarter-wave plate (QWP), a half-wave plate (HWP), and a polarising beam splitter (PBS). The pump beam ($\approx\SI{2.5}{\milli\watt}$) is reflected by gold-coated mirrors to preserve its polarization quality. The QWP rotates the pump beam’s polarization from vertical to circular (left- or right-handed), after which it is directed back into the vapor cell, ensuring good spatial overlap with the probe beam.

\begin{figure}[!ht]
    \centering
    \includegraphics[width=\columnwidth]{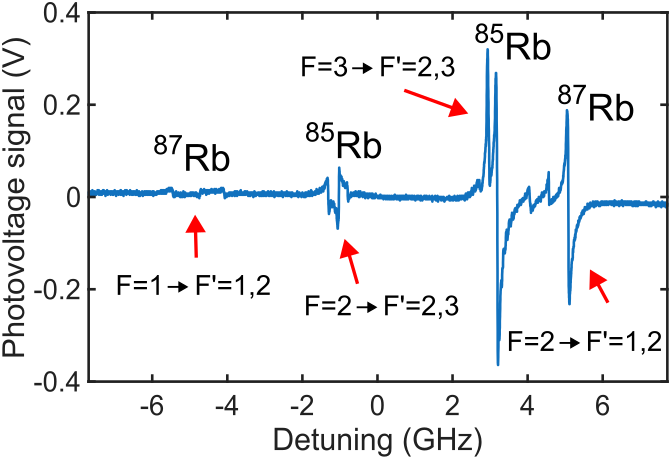}
    \caption{\textbf{Polarization spectroscopy spectrum.} The measured signal from balanced photodetectors on the frequency locking module while the frequency of the laser is scanned.}
    \label{fig:freqlocking}
\end{figure}

The probe beam ($\approx \SI{100}{\micro\watt}$) passes through an HWP to achieve linear polarisation. After transmission through the vapor cell, a second HWP is used to balance the power between the two photodiodes of the balanced detector preceded by a PBS. We noticed that the cancellation of the Doppler features from each photodiode while the pump is blocked can be optimised by adding a HWP between the PBS and the vapor cell. By adjusting the two HWP surrounding the vapor cell (corresponding to a walk-off alignment procedure), the cancellation of these Doppler features can be optimised, leading to a higher quality signal for frequency locking in the presence of the pump. Laser frequency locking is performed using a \textit{Moku:GO} device from \textit{Liquid Instruments}, enabling precise stabilisation of the laser to the desired atomic transition. By monitoring the locked signals, we reported few $\SI{}{\mega\hertz}$ stability over more than $\SI{1}{\hour}$ of acquisition time, which is sufficient for our needs. Further tuning of the locking parameters can provide more stable locks.

Fig.~\ref{fig:freqlocking} shows the hyperfine transitions for the $D_1$ line of $^{85}$Rb and $^{87}$Rb, while the frequency of the laser is linearly swiped. In this work, the vapor cell used to sense the magnetic fields is filled with $^{87}$Rb atoms. Therefore we locked the frequency of the laser onto the $^{87}$Rb hyperfine transition $F=2$ to $F'=2$.

\section{Calibration of the sample}
\label{Appendix:CalibSample}
The sample used in this work consists of 10 permanent magnets (axially magnetised, \textit{Part:9168}, \textit{Radial Magnets}) assembled together. 
\begin{figure}[!h]
    \centering
    \includegraphics[scale = 0.35]{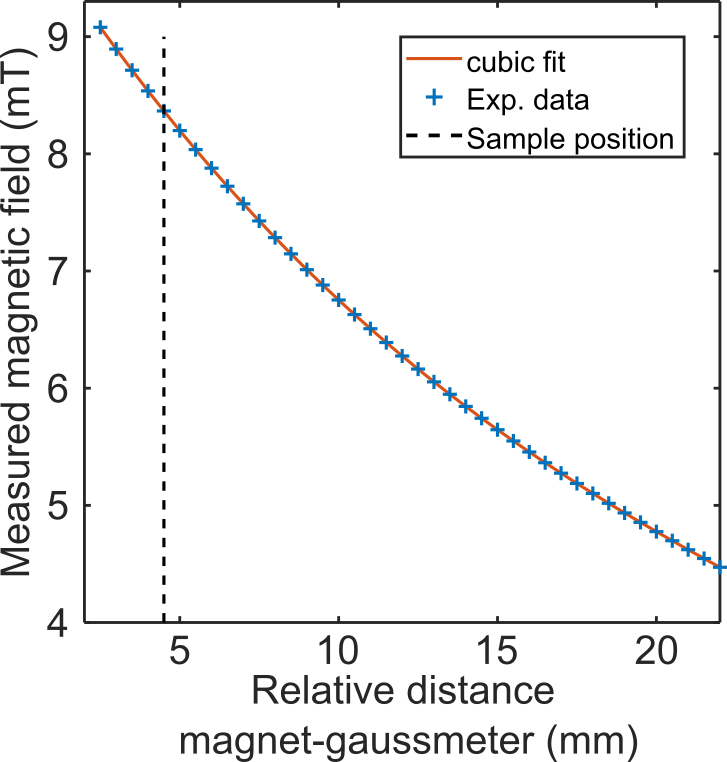}
    \caption{\textbf{Sample calibration.} The magnetic field measured by a gaussmeter while the sample is moved with motorised translation stage. The dashed line corresponds to the position of the translation stage during the magnetic field imaging illustrated in figure~\ref{fig:SPI_MF}.}
    \label{fig:samplecalib}
\end{figure}
It is placed on the left side of the cell when viewed along the $z$ direction (i.e., on the negative $x$ axis, as shown in Fig.~\ref{fig:SPI_MF}). To compare our SPI measurements with a reference, we used a gaussmeter positioned at the point where the light beam crosses the cell, aligned with the centre of the cell along the $z$ axis. Subsequently, the magnetic field along the direction of the light beam is measured while the sample is moved using a motorised translation stage (\textit{MTS25-Z8}, \textit{Thorlabs}).

Figure~\ref{fig:samplecalib} shows these measurements. The blue crosses represent the experimentally measured magnetic field along the $z$ axis as the sample was moved further away from the probe. The orange trace is a third-order polynomial fit, which matches the scaling predicted by a 3D COMSOL simulation. The dashed black line indicates the position of the translation stage during the magnetic field imaging shown in Fig.~\ref{fig:SPI_MF}.

While the method used to estimate the magnetic field gradient in the horizontal direction is precise (due to the motorized translation stage controlling the sample's movement), the relative distance between the probe and the sample was measured with a ruler, introducing an estimated uncertainty of $\pm \SI{2}{\mm}$. Since the spatial resolution of our magnetic field imaging setup is on the micrometre scale, shifting the sample by $\SI{2}{\mm}$ closer to or farther from the probe beam can noticeably affect the observed gradient of the magnetic field in the horizontal direction.

\bibliographystyle{unsrt}
\bibliography{MainPaper}

\begin{thebibliography}{10}

\bibitem{Kominis2003}
I.~K. Kominis, T.~W. Kornack, J.~C. Allred, and M.~V. Romalis.
\newblock A subfemtotesla multichannel atomic magnetometer.
\newblock {\em Nature}, 422(6932):596–599, 2003.

\bibitem{Dang2010}
H.~B. Dang, A.~C. Maloof, and M.~V. Romalis.
\newblock Ultrahigh sensitivity magnetic field and magnetization measurements with an atomic magnetometer.
\newblock {\em Applied Physics Letters}, 97(15), 2010.

\bibitem{Ledbetter2008}
M.~P. Ledbetter, I.~M. Savukov, V.~M. Acosta, D.~Budker, and M.~V. Romalis.
\newblock Spin-exchange-relaxation-free magnetometry with {Cs} vapor.
\newblock {\em Physical Review A}, 77(3), 2008.

\bibitem{BEC_MfieldImaging}
S.~Wildermuth, S.~Hofferberth, I.~Lesanovsky, E.~Haller, L.~M. Andersson, S.~Groth, I.~Bar-Joseph, P.~Kr{\"u}ger, and J.~Schmiedmayer.
\newblock {Bose-Einstein} condensates: microscopic magnetic-field imaging.
\newblock {\em Nature}, 435(7041):440, 2005.

\bibitem{Kitching2007}
V.~Shah, S.~Knappe, P.~D.~D. Schwindt, and J.~Kitching.
\newblock Subpicotesla atomic magnetometry with a microfabricated vapour cell.
\newblock {\em Nat. Photonics}, 1(11):649--652, 2007.

\bibitem{Compact_mag}
V.~Shah and M.~V. Romalis.
\newblock Spin-exchange relaxation-free magnetometry using elliptically polarized light.
\newblock {\em Phys. Rev. A}, 80(1), 2009.

\bibitem{Griffith2010}
W.~C. Griffith, S.~Knappe, and J.~Kitching.
\newblock Femtotesla atomic magnetometry in a microfabricated vapor cell.
\newblock {\em Opt. Express}, 18(26):27167--27172, 2010.

\bibitem{Fernholz2019}
T.~Pyragius, H.~M. Florez, and T.~Fernholz.
\newblock Voigt-effect-based three-dimensional vector magnetometer.
\newblock {\em Phys. Rev. A}, 100:023416, 2019.

\bibitem{Brookes2022}
M.~J. Brookes, J.~Leggett, M.~Rea, R.~M. Hill, N.~Holmes, El. Boto, and R.~Bowtell.
\newblock Magnetoencephalography with optically pumped magnetometers (opm-meg): the next generation of functional neuroimaging.
\newblock {\em Trends in Neurosciences}, 45(8):621–634, 2022.

\bibitem{Dawson2023}
R.~Dawson, Ca. O'Dwyer, M.~S. Mrozowski, E.~Irwin, J.~P. McGilligan, D.~P. Burt, D.~Hunter, S.~Ingleby, M.~Rea, N.~Holmes, M.~J. Brookes, P.~F. Griffin, and E.~Riis.
\newblock Portable single-beam cesium zero-field magnetometer for magnetocardiography.
\newblock {\em J. Optical Microsystems}, 3(04), 2023.

\bibitem{Westner2021}
B.~U. Westner, J.~I. Lubell, M.~Jensen, S.~Hokland, and S.~S. Dalal.
\newblock Contactless measurements of retinal activity using optically pumped magnetometers.
\newblock {\em NeuroImage}, 243:118528, 2021.

\bibitem{Narducci2009}
E.~E. Mikhailov, I.~Novikova, M.~D. Havey, and F.~A. Narducci.
\newblock Magnetic field imaging with atomic {Rb} vapor.
\newblock {\em Opt. Lett.}, 34(22):3529--3531, 2009.

\bibitem{Knappe2017}
O.~Alem, R.~Mhaskar, Ricardo J., D.~Sheng, J.~LeBlanc, L.~Trahms, T.~Sander, J.~Kitching, and S.~Knappe.
\newblock Magnetic field imaging with microfabricated optically-pumped magnetometers.
\newblock {\em Opt. Express}, 25(7):7849--7858, 2017.

\bibitem{Riis2023}
D.~Hunter, C.~Perrella, A.~McWilliam, J.~P. McGilligan, M.~Mrozowski, S.~J. Ingleby, P.~F. Griffin, D.~Burt, A.~N. Luiten, and E.~Riis.
\newblock Free-induction-decay magnetic field imaging with a microfabricated {Cs} vapor cell.
\newblock {\em Opt. Express}, 31(20):33582--33595, 2023.

\bibitem{Treutlein2012}
P.~B{\"o}hi and P.~Treutlein.
\newblock Simple microwave field imaging technique using hot atomic vapor cells.
\newblock {\em Appl. Phys. Lett.}, 101(18):181107, 2012.

\bibitem{Treutlein2015}
A.~Horsley, G.~Du, and P.~Treutlein.
\newblock Widefield microwave imaging in alkali vapor cells with sub-100$\mu$m resolution.
\newblock {\em New J. Phys.}, 17(11):112002, 2015.

\bibitem{Prajapati2024}
N.~Schlossberger, T.~McDonald, K.~Su, R.~Talashila, R.~Behary, C.~L. Patrick, D.~Hammerland, E.~E. Mikhailov, S.~Aubin, I.~Novikova, C.~L. Holloway, and N.~Prajapati.
\newblock Two-dimensional imaging of electromagnetic fields via light sheet fluorescence imaging with {R}ydberg atoms, 2024.

\bibitem{Mileti2015}
C.~Affolderbach, G.~Du, T.~Bandi, A.~Horsley, P.~Treutlein, and G.~Mileti.
\newblock Imaging microwave and {DC} magnetic fields in a vapor-cell {Rb} atomic clock.
\newblock {\em IEEE Trans. Instrum. Meas.}, 64(12):3629--3637, 2015.

\bibitem{Duarte2008}
M.~F. Duarte, M.~A. Davenport, D.~Takhar, J.~N. Laska, T.~Sun, K.~F. Kelly, and R.~G. Baraniuk.
\newblock Single-pixel imaging via compressive sampling.
\newblock {\em IEEE Signal Processing Magazine}, 25(2):83–91, 2008.

\bibitem{Schatz1969}
P.~N. Schatz and A.~J. McCaffery.
\newblock The {F}araday effect.
\newblock {\em Quarterly Reviews, Chemical Society}, 23(4):552, 1969.

\bibitem{Lancis2012}
V.~Dur{\'a}n, P.~Clemente, M.~Fern{\'a}ndez-Alonso, E.~Tajahuerce, and J.~Lancis.
\newblock Single-pixel polarimetric imaging.
\newblock {\em Opt. Lett.}, 37(5):824--826, 2012.

\bibitem{Padgett2015}
S.~S. Welsh, M.~P Edgar, R.~Bowman, B.~Sun, and M.~J. Padgett.
\newblock Near video-rate linear stokes imaging with single-pixel detectors.
\newblock {\em J. Opt.}, 17(2):025705, 2015.

\bibitem{Zhao2020}
H.~Wu, M.~Zhao, F.~Li, Z.~Tian, and M.~Zhao.
\newblock Underwater polarization‐based single pixel imaging.
\newblock {\em J. Soc. Inf. Disp.}, 28(2):157--163, 2020.

\bibitem{Foreman2020}
K.~L.~C. Seow, P.~T{\"o}r{\"o}k, and M.~R. Foreman.
\newblock Single pixel polarimetric imaging through scattering media.
\newblock {\em Opt. Lett.}, 45(20):5740--5743, 2020.

\bibitem{Wang2022}
Y.~Chen, K.~Yin, D.~Shi, W.~Yang, J.~Huang, Z.~Guo, K.~Yuan, and Y.~Wang.
\newblock Detection and imaging of distant targets by near-infrared polarization single-pixel lidar.
\newblock {\em Appl. Opt.}, 61(23):6905--6914, 2022.

\bibitem{Gibson2020}
G.~M. Gibson, S.~D. Johnson, and M.~J. Padgett.
\newblock Single-pixel imaging 12 years on: a review.
\newblock {\em Optics Express}, 28(19):28190, 2020.

\bibitem{Wolfgramm2010}
F.~Wolfgramm, A.~Cerè, F.~A. Beduini, A.~Predojević, M.~Koschorreck, and M.~W. Mitchell.
\newblock Squeezed-light optical magnetometry.
\newblock {\em Physical Review Letters}, 105(5), 2010.

\bibitem{Horrom2012}
T.~Horrom, R.~Singh, J.~P. Dowling, and E.~E. Mikhailov.
\newblock Quantum-enhanced magnetometer with low-frequency squeezing.
\newblock {\em Physical Review A}, 86(2), 2012.

\bibitem{Bai2021}
L.~Bai, X.~Wen, Y.~Yang, L.~Zhang, J.~He, Y.~Wang, and J.~Wang.
\newblock Quantum-enhanced rubidium atomic magnetometer based on {F}araday rotation via 795 nm {S}tokes operator squeezed light.
\newblock {\em Journal of Optics}, 23(8):085202, 2021.

\bibitem{SPIFvsH}
Z.~Zhang, X.~Wang, G.~Zheng, and J.~Zhong.
\newblock Hadamard single-pixel imaging versus {F}ourier single-pixel imaging.
\newblock {\em Opt. Express}, 25(16):19619, 2017.

\bibitem{freq_locking}
C.~P. Pearman, C.~S. Adams, S.~G. Cox, P.~F. Griffin, D.~A. Smith, and I.~G. Hughes.
\newblock Polarization spectroscopy of a closed atomic transition: applications to laser frequency locking.
\newblock {\em J. Phys. B At. Mol. Opt. Phys.}, 35(24):5141--5151, 2002.

\bibitem{Arduino_lib}
B.~Gombala.
\newblock {STMSpeeduino}, 2024.

\bibitem{Alexandrov2003-th}
E.~B. Alexandrov.
\newblock Recent progress in optically pumped magnetometers.
\newblock {\em Phys. Scr.}, T105(1):27, 2003.

\bibitem{Tierney2019}
T.~M. Tierney, N.~Holmes, S.~Mellor, Jos{\'e}~D. L{\'o}pez, G.~Roberts, R.~. Hill, E.~Boto, J.~Leggett, V.~Shah, M.~J. Brookes, R.~Bowtell, and G.~R. Barnes.
\newblock Optically pumped magnetometers: From quantum origins to multi-channel magnetoencephalography.
\newblock {\em Neuroimage}, 199:598--608, 2019.

\bibitem{Dai2018}
Y.~Zhang, J.~Suo, Y.~Wang, and Q.~Dai.
\newblock Doubling the pixel count limitation of single-pixel imaging via sinusoidal amplitude modulation.
\newblock {\em Opt. Express}, 26(6):6929, 2018.

\bibitem{Radwell2017}
M.~Sun, L.~Meng, M.~P. Edgar, M.~J. Padgett, and N.~Radwell.
\newblock A {R}ussian {D}olls ordering of the {H}adamard basis for compressive single-pixel imaging.
\newblock {\em Sci. Rep.}, 7(1):3464, 2017.

\bibitem{Liang2022}
P.~Kilcullen, T.~Ozaki, and J.~Liang.
\newblock Compressed ultrahigh-speed single-pixel imaging by swept aggregate patterns.
\newblock {\em Nat. Commun.}, 13(1):7879, 2022.

\bibitem{Mikhailov2008}
E.~E. Mikhailov and I.~Novikova.
\newblock Low-frequency vacuum squeezing via polarization self-rotation in {Rb} vapor.
\newblock {\em Optics Letters}, 33(11):1213, 2008.

\bibitem{Wen2017}
X.~Wen, Y.~Han, J.~Liu, J.~He, and J.~Wang.
\newblock Polarization squeezing at the audio frequency band for the {R}ubidium {$D_1$} line.
\newblock {\em Optics Express}, 25(17):20737, 2017.

\bibitem{Zhang2017}
M.~Zhang, M.~A. Guidry, R.~N. Lanning, Z.~Xiao, J.~P. Dowling, I.~Novikova, and E.~E. Mikhailov.
\newblock Multipass configuration for improved squeezed vacuum generation in hot {Rb} vapor.
\newblock {\em Physical Review A}, 96(1), 2017.

\bibitem{Zhao2025}
J.~Zhao, Z.~Yu, X.~Chen, Y.~Wu, X.~Liang, W.~Huang, K.~Zhang, C.~Yuan, and L.~Q. Chen.
\newblock Ultra-stable and high-performance squeezed vacuum source enabled via artificial intelligence control.
\newblock {\em Science Advances}, 11(18), 2025.

\bibitem{Taylor2013}
M.~A. Taylor, J.~Janousek, V.~Daria, J.~Knittel, B.~Hage, H.~Bachor, and W.~P. Bowen.
\newblock Biological measurement beyond the quantum limit.
\newblock {\em Nature Photonics}, 7(3):229–233, 2013.

\bibitem{SabinesChesterking2019}
J.~Sabines-Chesterking, A.~R. McMillan, P.~A. Moreau, S.~K. Joshi, S.~Knauer, E.~Johnston, J.~G. Rarity, and J.~C.~F. Matthews.
\newblock Twin-beam sub-shot-noise raster-scanning microscope.
\newblock {\em Optics Express}, 27(21):30810, 2019.

\bibitem{Atkinson2021}
G.S. Atkinson, E.J. Allen, G.~Ferranti, A.R. McMillan, and J.C.F. Matthews.
\newblock Quantum enhanced precision estimation of transmission with bright squeezed light.
\newblock {\em Physical Review Applied}, 16(4), 2021.

\bibitem{Pooser2013}
B.~J. Lawrie and R.~C. Pooser.
\newblock Toward real-time quantum imaging with a single pixel camera.
\newblock {\em Opt. Express}, 21(6):7549--7559, 2013.

\end{thebibliography}

\end{document}